\documentclass[preprint,showpacs]{revtex4}
\usepackage{amsmath}
\usepackage{graphicx}

\begin{document}

\title{Gluon Fusion induced $Zg$ and $Zgg$ Productions in the Standard Model at the LHC}
\author{Gao Xiangdong}
\address{Institute of High Energy Physics and Theoretical Physics Center for Science Facilities,
Chinese Academy of Sciences, Beijing 100049, People's Republic of China}
\author{Qiang Li}
\address{School of Physics, and State Key Laboratory of Nuclear Physics and Technology, Peking University, China}
\author{Cai-Dian L\"{u}}
\address{Institute of High Energy Physics and Theoretical Physics Center for Science Facilities,
Chinese Academy of Sciences, Beijing 100049, People's Republic of China}
\date{\today}

\begin{abstract}
We report calculations of the gluon induced $Zg$ and $Zgg$ productions in the Standard Model at the LHC operating at both 7 TeV and 14 TeV collision energy. We present total cross sections and differential distributions of the processes and compare them with the leading and next-to-leading order QCD $pp \rightarrow Z+1$ jet, $Z+2$ jets results. Our results show that the gluon induced $Zg$ and $Zgg$ productions contribute to $pp \rightarrow Z+1$ jet, $Z+2$ jets at $1\%$ level.
\end{abstract}

\maketitle

\section{Introduction}
The CERN Large Hadron Collider(LHC) is acquiring data at a high rate with the center mass energy of 7 TeV, which provides strong experimental support to explore the details of the Standard Model (SM) and investigate new physic (NP). One of the most important relevant processes at the LHC is $Z$ boson production accompanying jets($j$) as it is important for both quantitative tests of the SM and serving as backgrounds of NP processes. Thus it is necessary to estimate these processes as more accurate as possible. $Z +$ jets productions at hadron colliders have been extensively studied at QCD leading order(LO) and next-to-leading order(NLO)\cite{Berends:1989cf, Giele:1993dj, Campbell:2002tg, Campbell:2003hd, Berger:2010vm, Ita:2011wn} (and references therein).

$gg \rightarrow Zg$ and $gg \rightarrow Zgg$, although formally are finite subsets of the next-to-next-to-leading-order(NNLO) QCD corrections to the processes $pp \rightarrow Zj$ and $pp \rightarrow Zjj$, respectively, can get enhanced at the LHC because of high luminosity of gluon than quarks to compensate the suppression of $\alpha_s$ to some extent. Therefore, it is worthwhile to investigate these processes so as to judge whether their contributions could be ignored or not.
In Ref.~\cite{vanderBij:1988ac}, attention was paid to gluon induced $Z$ boson plus one jet associated production which however shows that the contribution is small as $\sim 1\%$ of the LO result.

In this paper, we update the calculations of $gg \rightarrow Zg$ and report also the results of $gg \rightarrow Zgg$ at the LHC. Comparing them with the LO and NLO $Zj$ and $Zjj$ production rates, we give an estimation of the importance of gluon induced $Zg$ and $Zgg$ productions at the LHC running at both 7 TeV and 14 TeV. This paper is organized as follows. In Sec. II, we give a brief description of our calculations. In Sec. III, we show numerical results and discussions. We close this paper with a summary.

\section{Description of the Calculation}

\begin{figure}[h]
  \centering
   \includegraphics[width=0.7\textwidth]{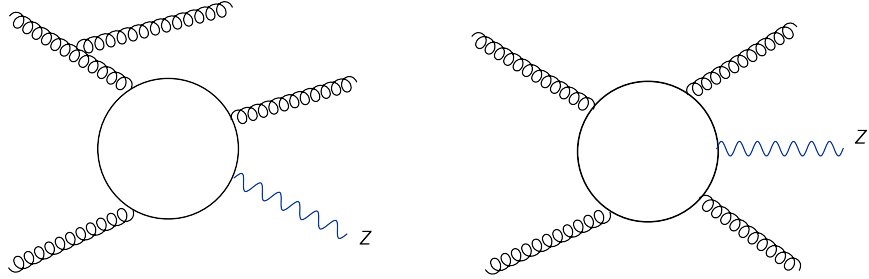}
    \caption{Feynman Diagrams generated by Jaxodraw~\cite{Vermaseren:1994je, Binosi:2003yf, Binosi:2008ig} for the partonic process $gg\rightarrow Z gg$, sorted into 2 topological classes. Taking into account of all possible permutation, one gets 18 diagrams for (a) and 12 ones for (b). In addition, one needs to sum over the Fermion flavors and flow directions within the Fermion loop.}
  \label{fd}
\end{figure}

The relevant one-loop Feynman diagrams and amplitudes for the partonic process $gg\rightarrow Z gg$
have been generated with FeynArts 3.6~\cite{Kublbeck:1990xc, Hahn:2000kx}. The diagrams are sorted
into 2 topological classes, corresponding to boxes and pentagons,
as shown in Fig.~\ref{fd}. Note that the contributions of triangle diagrams can be neglected~\cite{vanderBij:1988ac}.

We then use FormCalc 6.2 \cite{Hahn:1998yk} to manipulate the amplitudes. The Fortran libraries generated with the FormCalc are then linked to our own Monte Carlo integration code for numerical calculations. The tensor integrals are evaluated with the LoopTools-2.5 \cite{Hahn:1998yk}, which uses the method introduced in Ref.~\cite{Denner:2002ii} to reduce the pentagon tensor integrals, and Passarino-Veltman procedure \cite{Passarino:1978jh, Consoli:1979xw, Veltman:1980fk, Green:1980bd}to reduce lower point ones. The resulting regular scalar integrals are evaluated with the FF package \cite{vanOldenborgh:1989wn, vanOldenborgh:1990yc}. Moreover, we have implemented the reduction method for pentagon tensor integrals up to rank 5 as proposed in Ref.~\cite{Denner:2005nn}. We have also implemented in LoopTools the so called alternative Passarino-Veltman reduction for triangle and box tensor integrals~\cite{Denner:2005nn} to improve numerical stability. We use MCFM \cite{Campbell:2002tg, Campbell:2003hd} to generate results of the $pp \rightarrow Zj, Zjj$ to make a comparison with our results.

We checked the cancelation of ultraviolet and infrared divergences in our calculations, and tested the independence on the mass scale in the Fermion loop. We also checked our results by comparing with numerical results shown in Ref. \cite{vanderBij:1988ac}. We reproduced their results with the same settings and parameters.
\section{Numerical Results}
In this section we present our numerical results for the total and differential cross sections of the $Z+j$ and $Z+jj$ productions at the LHC. We use CTEQ6L1/CTEQ6M parton distribution functions(PDFs)\cite{Pumplin:2002vw} and default values of corresponding LHAPDF\cite{Whalley:2005nh} strong coupling $\alpha_s$ for the calculations of the LO/NLO subprocesses. The masses of the $W$ and $Z$ boson and the Fermi constant are taken as input, while the Weinberg angle and the electromagnetic coupling are then calculated using tree level relations of the SM parameters. Hence, numerically the SM parameters used in our calculations are listed below:
\begin{eqnarray}
&m_W = 80.398\text{GeV},~~m_Z = 91.1876 \text{GeV},~~G_F = 1.16637^{-5} \text{GeV}^{-2},&\nonumber\\
&\alpha(m_Z) = 1/132.3407,~~\sin^2\theta_W = 0.22264&
\end{eqnarray}
We impose the following cuts to define the jets
\begin{equation}
P_T^j>40 \text{GeV},~~|\eta_j|<4.5,~~\Delta R_{jj}>0.6,
\end{equation}
where $P_T^j$ is the transverse momentum of the jets, $\eta_j$ is the pseudorapidity of the jets
and $\phi$ is the azimuthal angle around the beam direction. $\Delta R_{jj}$ define the separation of two jets.
Renormalization and factorization scales are set as follows:
\begin{equation}
\mu_r = \mu_f = \mu_0\equiv\sqrt{P_{TZ}^2+m_Z^2}.
\end{equation}
Throughout the calculations, six flavors of quark are included in the quark loop with $m_b=4.6\,$GeV and $m_t=173\,$GeV, and all other quarks are taken as massless.

\begin{figure}[h]
\includegraphics[scale=0.9]{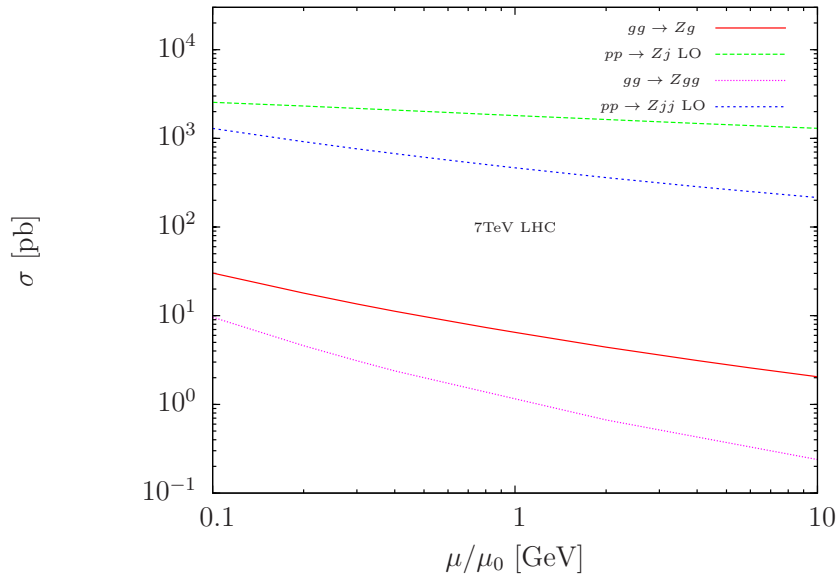}
\caption{\label{sig7}Scale dependences of the total cross sections for $gg\to Zg\,, Zgg$ and the LO $pp\to Zj,\,Zjj$ at the LHC with $\sqrt{s} = 7$ TeV.}
\end{figure}

\begin{figure}[h]
\includegraphics[scale=0.9]{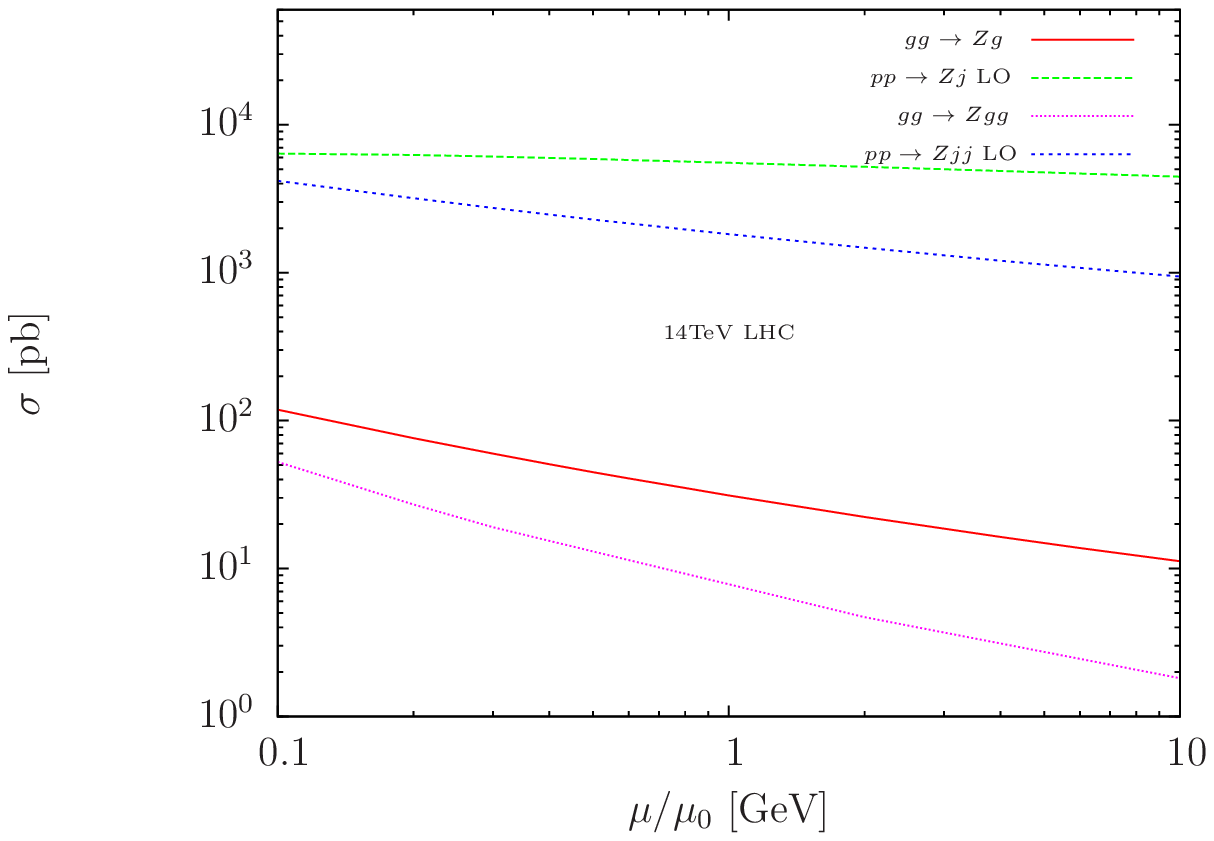}
\caption{\label{sig14}Scale dependences of the total cross sections for $gg\to Zg\,, Zgg$ and the LO $pp\to Zj,\,Zjj$  at the LHC with $\sqrt{s} = 14$ TeV.}
\end{figure}

Fig.~\ref{sig7} and Fig.~\ref{sig14} show the renormalization and factorization scales dependences of the total cross sections of gluon fusion induced $Zg$ and $Zgg$ productions and LO $pp \rightarrow Zj, Zjj$ processes at the LHC with $\sqrt{s} = 7$ TeV and 14 TeV, respectively. It can be read from Fig.~\ref{sig7} that at the 7 TeV LHC, the total cross section of the gluon induced $Zg$ production varies from 2.1pb to 30.3pb, while $Zgg$ production is from 0.24pb to 9.6pb. Both of the two processes contribute small rates to the $Z + $jet(s) processes at large scale, however, they counts almost $1\%$ of the total cross sections of the $Z + $jet(s) productions at small scale ($\sim 0.1 \mu_0$), as expected due to larger gluon flux there. At the 14 TeV LHC as in Fig.~\ref{sig14}, the total cross sections of the $Zg$ and $Zgg$ productions are around $10\sim120$ pb and $2\sim50$ pb, which is almost $2\%$ and $1.5\%$ of the total cross sections of the $Z + $jet(s) productions at small scale ($\sim 0.1 \mu_0$). The scale dependence of the gluon induced $Zg$ and $Zgg$ productions in both Fig.~\ref{sig7} and Fig.~\ref{sig14} are larger comparing with LO $Z + $jet(s) productions for the reason that the loop induced results with higher power of $\alpha_s$ shown here are yet partonic LO.

\begin{figure}[h]
\includegraphics[scale=0.9]{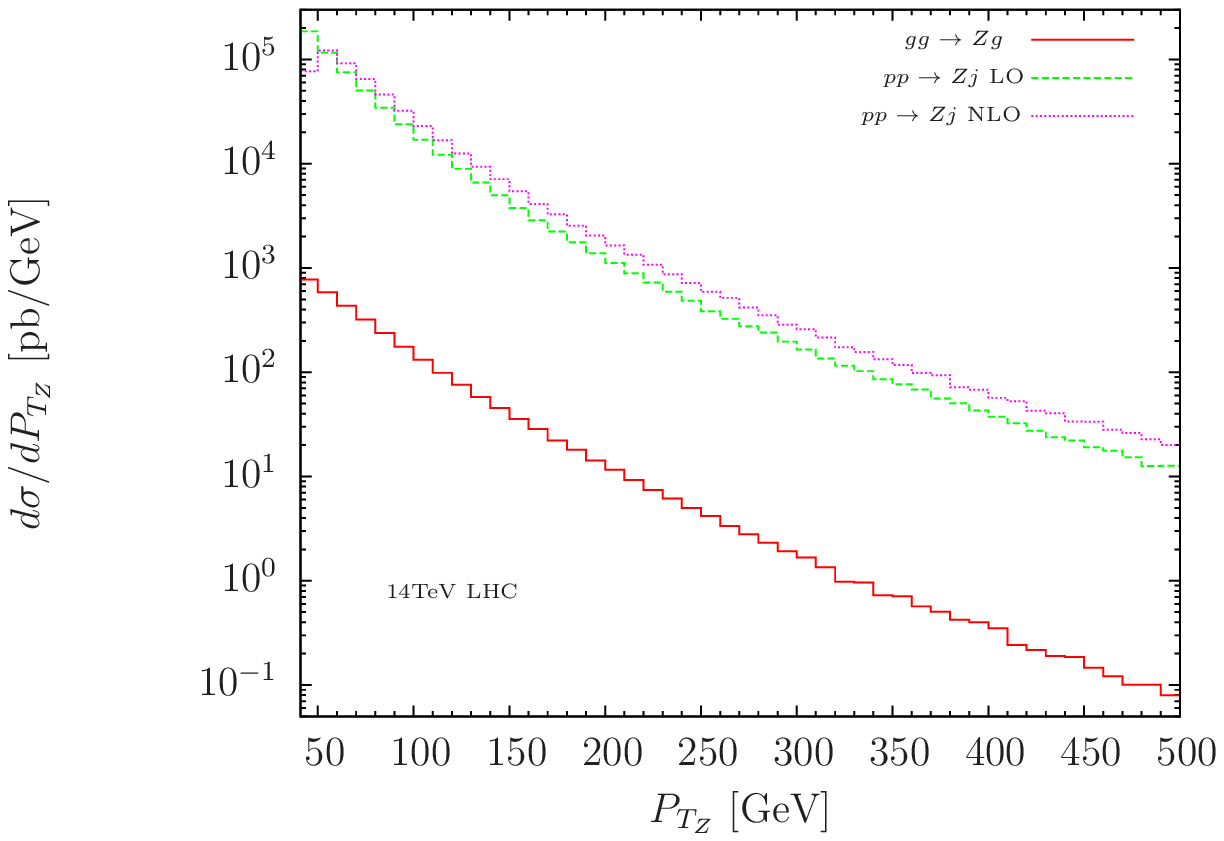}
\caption{\label{ptz1}Differential distributions of the transverse momentum of the $Z$ boson for the process $gg \rightarrow Zg$ and $pp \rightarrow Zj$ at the LHC with with $\sqrt{s} = 14$ TeV. The corresponding NLO curves of $pp \rightarrow Zj$ is also plotted with the help of MCFM. }
\end{figure}

\begin{figure}[h]
\includegraphics[scale=0.9]{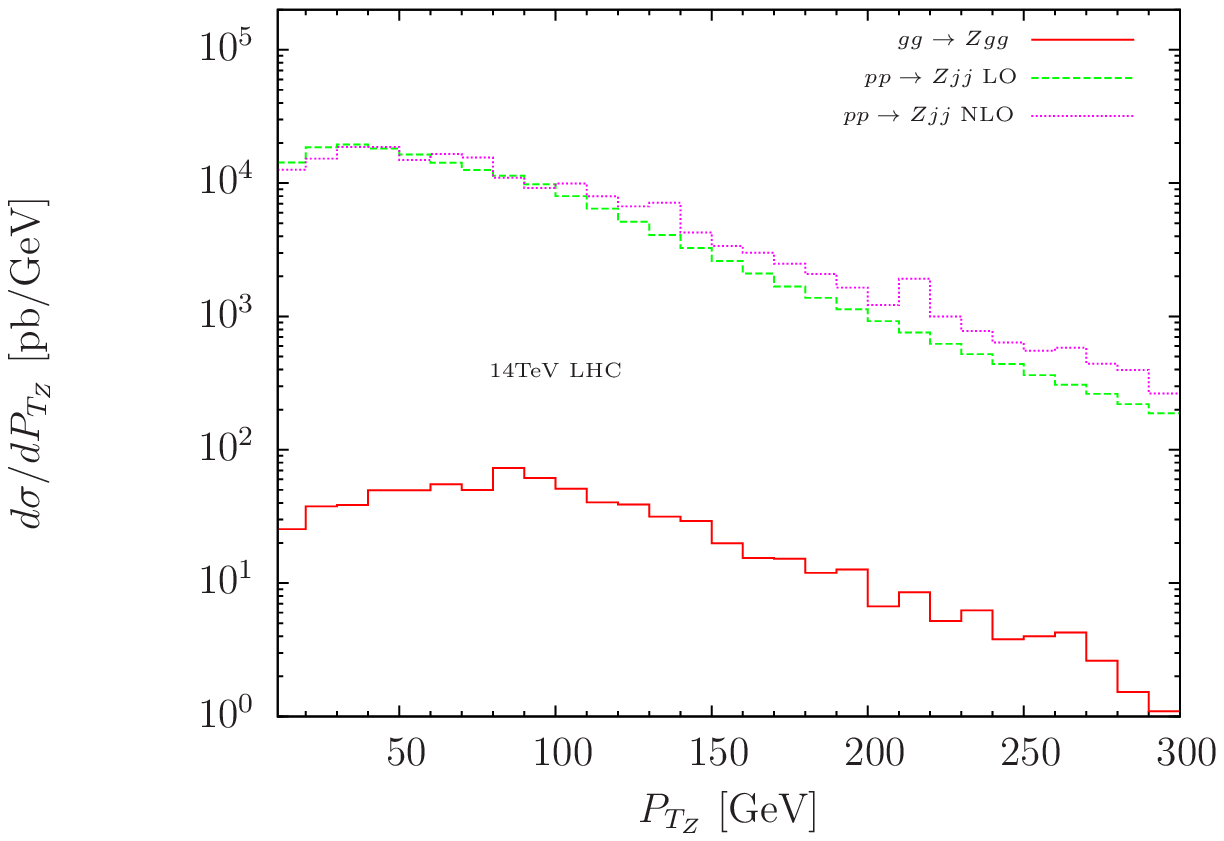}
\caption{\label{ptz2}Differential distributions of the transverse momentum of the $Z$ boson for the process $gg \rightarrow Zgg$ and $pp \rightarrow Zjj$ at the LHC with with $\sqrt{s} = 14$ TeV. The corresponding NLO curves of $pp \rightarrow Zjj$ is also plotted with the help of MCFM.}
\end{figure}

Fig.~\ref{ptz1} and Fig.~\ref{ptz2} display the differential distributions of the transverse momentum of the $Z$ boson for the gluon induced $Zg$ and $Zgg$ productions as well as the corresponding LO and NLO $Z +$jet(s) productions at the LHC with $\sqrt{s} = 14$ TeV. It can be seen from the two figures that the gluon induced $Zg$ and $Zgg$ productions do not change the shape of the differential distributions of the LO and NLO $Z +$jet(s) productions, and contribute to the $Z +$jet(s) productions at a rate similar to that in the total cross sections.

\section{Summary}
We have calculated the $Zg$ and $Zgg$ production processes induced by gluon fusion in the framework of the SM at the LHC. Our results show that the total cross sections of the $Zg$ and $Zgg$ productions process at the LHC can be around at $2\sim30$ pb with $\sqrt{s} = 7$ TeV and $10\sim100$ pb with $\sqrt{s} = 14$ TeV, and $0.2\sim10$ pb with $\sqrt{s} = 7$ TeV and $2\sim50$ pb with $\sqrt{s} = 14$ TeV, respectively. We also present the differential distributions of the transverse momentum of the $Z$ boson for the two processes. By comparing our results with $pp \rightarrow Zj$ and $pp \rightarrow Zjj$, we conclude that these two channels contribute at an order of $1\sim2\%$ to the $Zj$ and $Zjj$ productions at the LHC, at both the total and differential cross section levels.

\section*{Acknowledgements}
This work is partially supported by National Natural Science Foundation of China under
the Grant No. 10735080, and 11075168; Natural Science Foundation of Zhejiang Province of
China, Grant No. Y606252 and Scientific Research Fund of Zhejiang Provincial Education
Department of China, Grant No. 20051357; and China Postdoctoral Science Foundation
under Grant No. 20100480466.

\bibliography{Zjets}

\end{document}